# Anomalous impact of hydrostatic pressure on superconductivity of polycrystalline $LaO_{0.5}F_{0.5}BiSe_2$

Rajveer Jha and V.P.S. Awana*

*CSIR-National Physical Laboratory, Dr. K. S. Krishnan Marg, New Delhi-110012, India*

We report bulk superconductivity at 2.5K in $LaO_{0.5}F_{0.5}BiSe_2$ compound through the DC magnetic susceptibility and electrical resistivity measurements. The synthesized $LaO_{0.5}F_{0.5}BiSe_2$ compound is crystallized in tetragonal structure with space group P4/nmm and Reitveld refined lattice parameters are a= 4.15(1)Å and c=14.02(2)Å. The lower critical field of $H_{c1}$= 40Oe, at temperature 2K is estimated through the low field magnetization measurements. The $LaO_{0.5}F_{0.5}BiSe_2$ compound showed metallic normal state electrical resistivity with residual resistivity value of 1.35mΩ-cm. The compound is type-II superconductor, and the estimated $H_{c2}(0)$ value obtained by WHH formula is above 20kOe for 90%$\rho_n$ criteria. The superconducting transition temperature decreases with applied pressure till around 1.68GPa and with further higher pressures a high $T_c$ phase emerges with possible onset $T_c$ of above 5K for 2.5GPa.

***Corresponding Author**
Dr. V. P. S. Awana,
Principal Scientist
E-mail: awana@mail.npindia.org
Ph. +91-11-45609357, Fax-+91-11-45609310
Homepage awanavps.wenbs.com

## INTRODUCTION

The recent discovery of $BiS_2$ based materials with layered structure have been studied in past couple of years as a potential approach to the exploration of new superconductors. The novel $BiS_2$ based superconductivity has been reported first in $Bi_4O_4S_3$ layered compound with $T_c$ value of 4.5K [1, 2]. $Bi_4O_4S_3$ is composed of $Bi_2O_2(SO4)_{(1-x)}$ (x=0.5) blocking layers and $BiS_2$ superconducting layers [1]. Consequently, other $BiS_2$-based superconductors were discovered, including $ReO_{1-x}F_xBiS_2$ (Re= La, Ce, Nd, Yb, Pr), $La_{1-x}M_xOBiS_2$ (M= Ti, Zr, Hf, Th), and $Sr_{1-x}La_xFBiS_2$ [3-10]. The highest $T_c$=10.6 K has been obtained for $LaO_{0.5}F_{0.5}BiS_2$ superconductors under hydrostatic pressure [11]. In particular, the $T_c$ of $REO_{1-x}F_xBiS_2$ (RE=La, Ce, Nd, Pr) [11–14] and $Sr_{1-x}RE_xFBiS_2$ (RE = La, Ce, Nd, Pr, Sm) systems, enhances tremendously by applying the hydrostatic pressure [15, 16]. As far as the chemical substitutions are concerned the doping of Se on the S site in the $Bi_4O_4S_{3-x}Se_x$ and $NdOFBiS_{2-x}Se_x$ superconductors showed the suppression in $T_c$ [17, 18]. Interestingly, the superconductivity at 2.5K in $BiSe_2$ based layered $LaO_{0.5}F_{0.5}BiSe_2$ polycrystalline compound has recently been reported by A. Krzton-Maziopa, at. el, [19]. The crystal structure of the superconducting oxy-selenide is similar to other layered superconducting compounds such as CuO based cuprates and iron-based Pnictides with alternating stacks of blocking and superconducting layers [19]. Soon after this report, superconductivity with $T_c$ value of 3.0K was reported in $LaO_{0.5}F_{0.5}BiSe_2$ single crystals [20]. Also, the theoretical approaches were applied, and it was shown that the electronic structure, lattice dynamics, and electron–phonon interaction of the newly discovered superconductor $LaO_{0.5}F_{0.5}BiSe_2$ is that same as that for $LaO_{0.5}F_{0.5}BiS_2$ compound [21]. The pressure effect on $LaO_{0.5}F_{0.5}BiSe_2$ single crystals showed a decrease in $T_c$ with applied pressure of up to 1.75GPa and obtained high $T_c$ phase (6K) for higher pressure (> 1.97GPa) [22]. This is strikingly different than the impact of hydrostatic pressure on similar structure oxy-sulfides, where $T_c$ increases profoundly with pressure [11-16]. The external pressure effect is a conventional method for structural modulation that is often used to change the superconducting $T_c$. For example, the increase in $T_c$ of $LaO_{0.5}F_{0.5}BiS_2$ compound under pressure is due to a structural phase transition from tetragonal phase (P4/nmm) to monoclinic phase (P21/m) [23].

Here we report the successful synthesis of polycrystalline $LaO_{0.5}F_{0.5}BiSe_2$ compound through the solid state reaction method. The compound is crystallized in the tetragonal structure with space group P4/nmm. The bulk superconductivity at 2.5K has been observed through DC magnetic and electrical resistivity measurements. Because a positive pressure effect on $T_c$ has also been reported for $BiS_2$-based polycrystalline superconducting compounds [11-16] therefore we decided to determine the hydrostatic pressure effect on $T_c$ of polycrystalline $LaO_{0.5}F_{0.5}BiSe_2$ compound. We applied hydrostatic pressure on the polycrystalline $LaO_{0.5}F_{0.5}BiSe_2$ compound, and found that $T_c$ decreases with increasing applied pressure till around 1.68GPa and a high $T_c$ phase emerges at higher applied pressure of 2.5GPa with possible onset $T_c$ at around 5K. It is interesting to note that the impact of pressure on superconductivity of currently studied $BiSe_2$ is dramatically different to that as for $BiS_2$ ones [11-16]. Our results are qualitatively similar with the only available recent report [22] on impact of pressure on superconductivity of single crystalline $LaO_{0.5}F_{0.5}BiSe_2$.

## EXPERIMENTAL DETAILS

The polycrystalline bulk $LaO_{0.5}F_{0.5}BiSe_2$ sample was synthesized by standard solid state reaction route via vacuum encapsulation. High purity (4N) La, Bi, Se, $LaF_3$, and $La_2O_3$ are weighed in stoichiometric ratio and ground thoroughly in a glove box in high purity argon atmosphere. The mixed powders are subsequently palletized and vacuum-sealed ($10^{-4}$ mbar) in a quartz tube. The box furnace has been used to sinter the samples at $780^0$C for 12h with the typical heating rate of $2^o$C/min. The sintered sample is subsequently cooled down slowly to room temperature. This process has been repeated two times. X-ray diffraction (XRD) was performed at room temperature in the scattering angular ($2\theta$) range of $10^o$-$80^o$ in equal $2\theta$ step of $0.02^o$ using *Rigaku Diffractometer* with $Cu\ K_\alpha$ ($\lambda$ = 1.54Å). Rietveld analysis was performed using the standard *FullProf* program. DC magnetic susceptibility and transport measurements were performed on Physical Property Measurements System (*PPMS*-140kOe, *Quantum Design*).

The electrical resistivity under hydrostatic pressure measurements were carried out on Physical Property Measurements System (*PPMS*-140kOe, *Quantum Design*) by using HPC-33 Piston type pressure cell. Hydrostatic pressures were generated by a BeCu/NiCrAl clamped piston-cylinder cell. The sample was immersed in a fluid (Daphne Oil) pressure transmitting medium in a Teflon cell. Annealed Pt wires were affixed to gold-sputtered contact surfaces on each sample with silver epoxy in a standard four-wire configuration.

## RESULTS AND DISCUSSION

Figure 1(a) demonstrates the room temperature observed and Rietveld fitted XRD pattern of polycrystalline $LaO_{0.5}F_{0.5}BiSe_2$ compound. The compound is crystallized in tetragonal structure with the space group P4/nmm. Lattice parameters being obtained from Rietveld refinement of XRD are a= 4.15(1)Å and c=14.02(2)Å, which are in good agreement with earlier reports on $LaO_{0.5}F_{0.5}BiSe_2$ compound [19]. Small impurity peak of $Bi_2Se_3$ has also been seen and is marked as (*) in Figure 1, which is close to the background of the XRD pattern. The unit cell of the $LaO_{0.5}F_{0.5}BiSe_2$ compound is shown in figure 1(b). The compound has layered structure; including $(LaO)_2$ (Rare earth oxide) and $BiSe_2$ layers. Different atoms like Bismuth (Bi), Lanthanum (La), and Selenium (Se1 and Se2) occupy the 2c (0.25, 0.25, z) site, while the O/F atoms are at 2a (0.75, 0.25, 0) site. Interestingly, the carrier doping mechanism is same as Iron based compound LaFeAsO/F [24], where the carriers are doped from LaO/F to superconducting FeAs layer. Similarly, in $LaO_{0.5}F_{0.5}BiSe_2$, mobile carriers are doped from LaO/F redox layer to superconducting $BiSe_2$ layer.

Figure 2 depict the temperature dependent dc magnetic susceptibility measurements in Zero Field Cool (ZFC) and Field Cool (FC) protocols for the $LaO_{0.5}F_{0.5}BiSe_2$ compound under 10Oe magnetic field. The bulk superconductivity is confirmed as diamagnetic signal below superconducting transition with an onset $T_c$ at 2.5K, whereas the field-cooled susceptibility exhibits only a small drop, possibly due to a strong flux pinning effect. Inset of the figure 2 illustrates the lower critical field $H_{c1}(T)$ in low field isothermal magnetization (MH) measurements. The initial flux penetration and the deviation from linearity within the Meissner state defines the $H_{c1}(T)$. The lower critical field of studied $LaO_{0.5}F_{0.5}BiSe_2$ compound is 40Oe at temperature 2K.

Figure 3(a) shows the temperature dependence of electrical resistivity for $LaO_{0.5}F_{0.5}BiSe_2$ compound in the temperature range 2K and 300K. The normal state electrical resistivity behavior is metallic and comparable to the $Bi_4O_4S_3$ superconductor [1, 2]. The metallic behavior of normal state electrical resistivity has also been seen in $BiS_2$ based polycrystalline $Sr_{1-x}RE_xFBiS_2$ (RE= La, Ce, Nd, Pr, Sm) systems but under high pressure measurement [15, 16]. Interestingly, the $LaO_{0.5}F_{0.5}BiSe_2$ compound shows the metallic behavior at ambient pressure itself, which is well fitted to the resistivity linear equation $\rho=\rho_o+ AT$, where $\rho_o$ is the residual resistivity and A is the slope of the graph. Fitting of the $\rho(T)$ graph shows as red line in the temperature range 300K-50K in figure 3a and obtained residual resistivity value is 1.35mΩ-cm. The obtained value of residual resistivity is slightly higher than the one as reported for $LaO_{0.5}F_{0.5}BiSe_2$ compound earlier [19], suggesting possible inhomogeneity in our sample. The superconducting transition is clearly seen below $T_c$ onset of around 2.8 K, see inset of figure 3. The normal state semiconducting behavior of similar superconductor $REO_{0.5}F_{0.5}BiS_2$ compound has earlier been seen in many reports [3-10], which is quite different from metallic character of resistivity of $LaO_{0.5}F_{0.5}BiSe_2$ compound. Inset of the figure 3 shows the electrical resistivity under various magnetic fields in superconducting transition temperature region. The $T_c$ onset and $T_c$ ($\rho$=0) of $LaO_{0.5}F_{0.5}BiSe_2$ compound shift towards the low temperature side with applied magnetic field. It is the clear indication of type-II superconductivity like high $T_c$ cuprate and Iron based Pnictides superconductors. Figure 4 shows the upper critical field ($H_{c2}$) corresponding to the temperatures where the resistivity drops to 90% of the normal state resistivity. The $H_{c2}(0)$ is estimated by using the conventional one-band Werthamer–Helfand–Hohenberg (WHH) equation, i.e., $H_{c2}(0)=-0.693T_c(dH_{c2}/dT)_{T=Tc}$. The solid line is the result of fitting of $H_{c2}(T)$ to the WHH formula. The estimated $H_{c2}(0)$ is above 20kOe for 90%, $\rho_n$ criteria.

Figure 5 shows the temperature dependence of resistivity for the polycrystalline $LaO_{0.5}F_{0.5}BiSe_2$ compound at various hydrostatic pressures with temperatures ranging from 2K to 250K. The normal state resistivity of compound is metallic at ambient pressure as well in hydrostatic pressure of up to 2.5GPa. Interestingly, the normal state resistivity first decreases with increasing pressure up to 1.97GPa and then slightly increases for 2.20Gpa and 2.50GPa with still having a metallic character. Inset of figure 5 shows enlarged views of the superconducting transition region in the temperature range 6-2K under various pressures for $LaO_{0.5}F_{0.5}BiSe_2$ compound. It can be clearly seen that the $T_c$ at ambient pressure gradually decreases with increasing pressure. Interestingly, a high $T_c$ phase emerges at applied pressure above 2GPa with maximum possible onset $T_c$ of around 5K for 2.5GPa pressure. Interestingly, we did not obtained $T_c(\rho=0)$ for the high $T_c$ (5K onset) phase. The Superconducting transition $T_c(\rho=0)$ may occur for the higher applied pressures as observed in a previous report for the same $LaO_{0.5}F_{0.5}BiSe_2$ single crystalline compound [22].

Interestingly, $T_c$ increases monotonically with applying pressure on $BiS_2$-based similar structure superconductors [11-16]. This is clear dissimilarity between the presently studied $BiSe_2$-based superconductors and the by now widely studied $BiS_2$-based ones, particularly the $REO_{1-x}F_xBiS_2$ and $Sr_{1-x}RE_xFBiS_2$ [11-16]. It has been suggested the increase in the superconducting $T_c$ with increasing pressure for polycrystalline $LaO_{0.5}F_{0.5}BiS_2$ compound is due to structural phase transition from a tetragonal phase (P4/nmm) to a monoclinic phase (P21/m) [23]. However, by examining available data (ref. 22, and the present results) on the relative pressure effect on the $T_c$ of $LaO_{0.5}F_{0.5}BiSe_2$ superconductor, clearly negative $dT_c/dP$ is seen for low pressures of up to say 2GPa, and later a high $T_c$ (5K) phase appears for higher pressures. The decrease in the superconducting transition with applied low applied pressures (<2GPa) seems to be due to possible over doping, as the normal state resistivity decreases in this regime. Later for higher pressures of above 2GPa, the high $T_c$ phase might be appearing with a new crystal structure as seen for $LaO_{0.5}F_{0.5}BiS_2$ compound [22]. The under pressure structural details at various temperatures for the studied $LaO_{0.5}F_{0.5}BiSe_2$ superconductor are strictly warranted, before commenting on real cause of the anomalous pressure dependence of superconducting transition temperature ($T_c$) of the same.

CONCLUSION:

In conclusion, we have successfully synthesized the polycrystalline $LaO_{0.5}F_{0.5}BiSe_2$ compound in single phase with superconductivity ($T_c$) at above 2.8K in both magnetic and electrical resistivity measurements. The estimated $H_{c2}(0)$ value of the compound is above 20kOe. The $T_c$ of the compound first decreases with applied pressure till around 1.68GPa and for higher pressures a high $T_c$ phase emerges with possible onset $T_c$ of above 5K for 2.5GPa. The impact of hydrostatic pressure on the presently studied $LaO_{0.5}F_{0.5}BiSe_2$ is much different than the widely studied $LaO_{0.5}F_{0.5}BiS_2$ superconductor. Structural details under pressure for $LaO_{0.5}F_{0.5}BiSe_2$ are much warranted.

ACKNOWLEDGEMENT:

Authors would thank their Director NPL-CSIR, India for his consistent support and interest in the present work. This work is financially supported by DAE-SRC outstanding investigator award scheme on search for new superconductors.

## Figure Captions

**Figure 1:** (a) Reitveld fitted room temperature XRD patterns for $LaO_{0.5}F_{0.5}BiSe_2$, compound (b) schematic unit cell of $LaO_{0.5}F_{0.5}BiSe_2$ compound.

**Figure 2:** Temperature dependence of DC Magnetic susceptibility in ZFC and FC modes for $LaO_{0.5}F_{0.5}BiSe_2$, compound at 10Oe, inset shows the low field magnetization to estimate lower critical field.

**Figure 3:** Temperature dependence of electrical resistivity ρ (T) plot for $LaO_{0.5}F_{0.5}BiSe_2$ compound in the temperature range 300K-2.0 K. Inset show the ρ (T)H plot at various magnetic fields in the temperature range 4-2K.

**Figure 4:** The upper critical field $H_{c2}(0)$ Vs T plots of the $LaO_{0.5}F_{0.5}BiSe_2$ compound for 90%, $\rho_n$ criteria.

**Figure 5:** ρ Vs T plots for $LaO_{0.5}F_{0.5}BiSe_2$ compound, at varying pressures in the temperature range 300K-2.0 K. Inset is ρ Vs T plots for $LaO_{0.5}F_{0.5}BiSe_2$ compound, at varying pressures in the temperature range 6K-2.0 K.

## FIGURES:

Fig. 1(a)

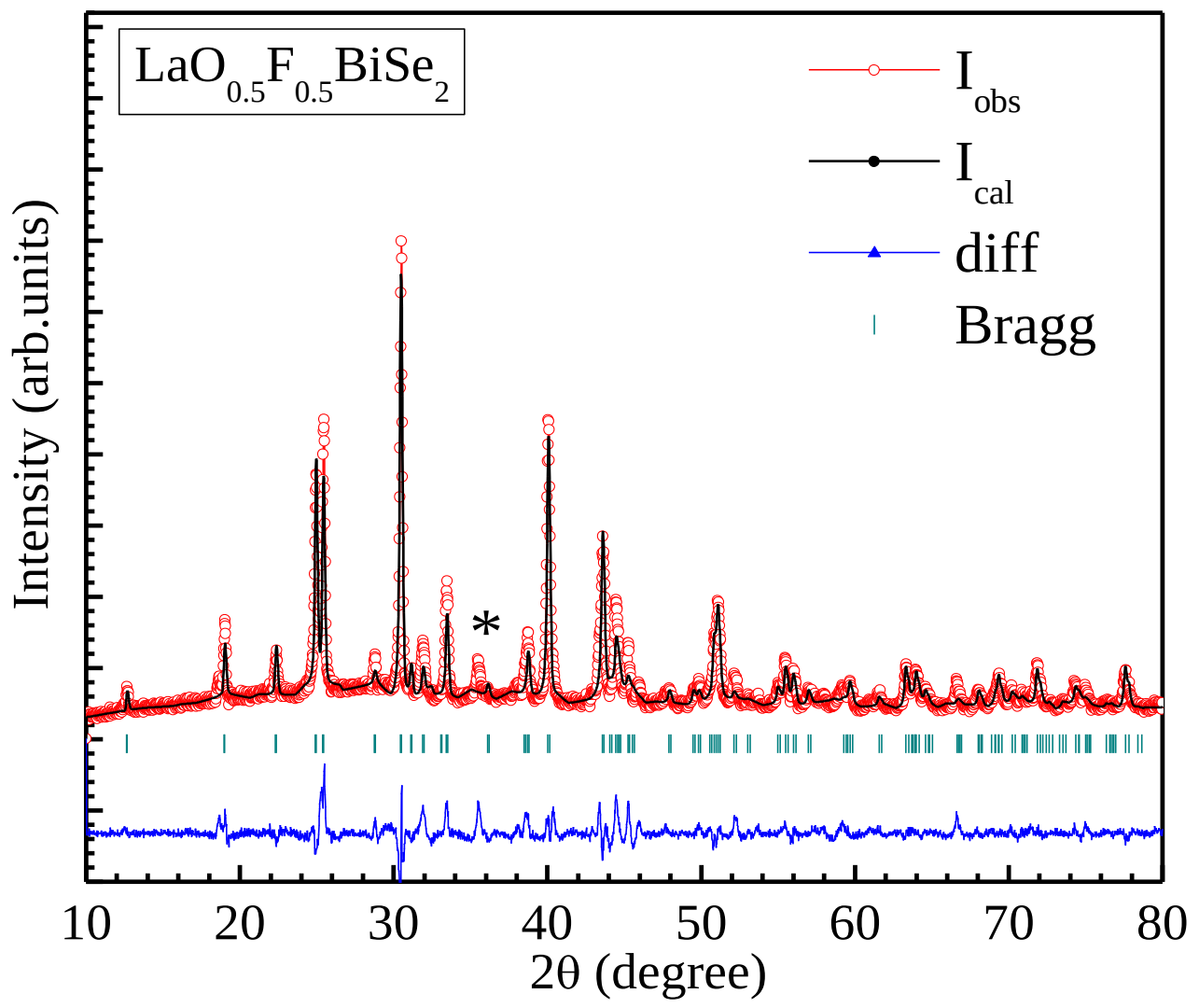

Fig. 1(b)

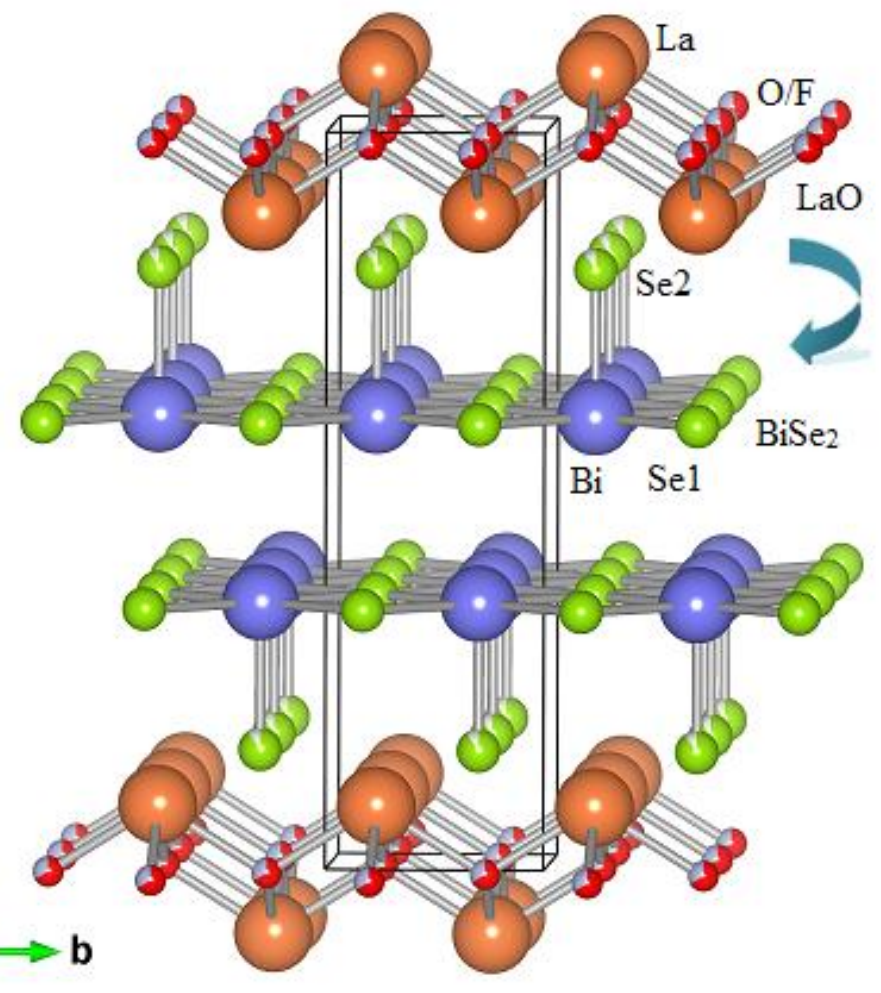

Fig. 2

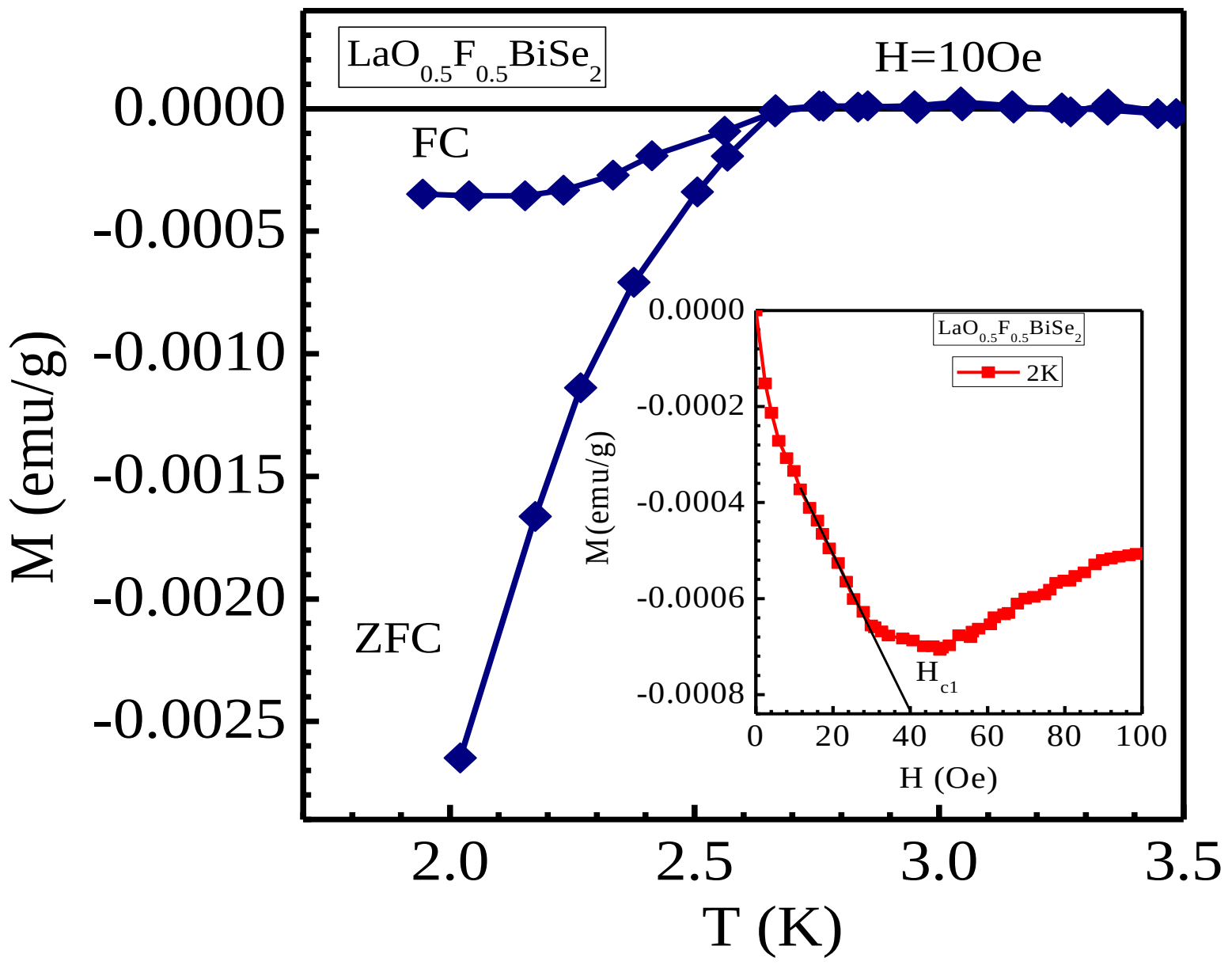

Fig. 3

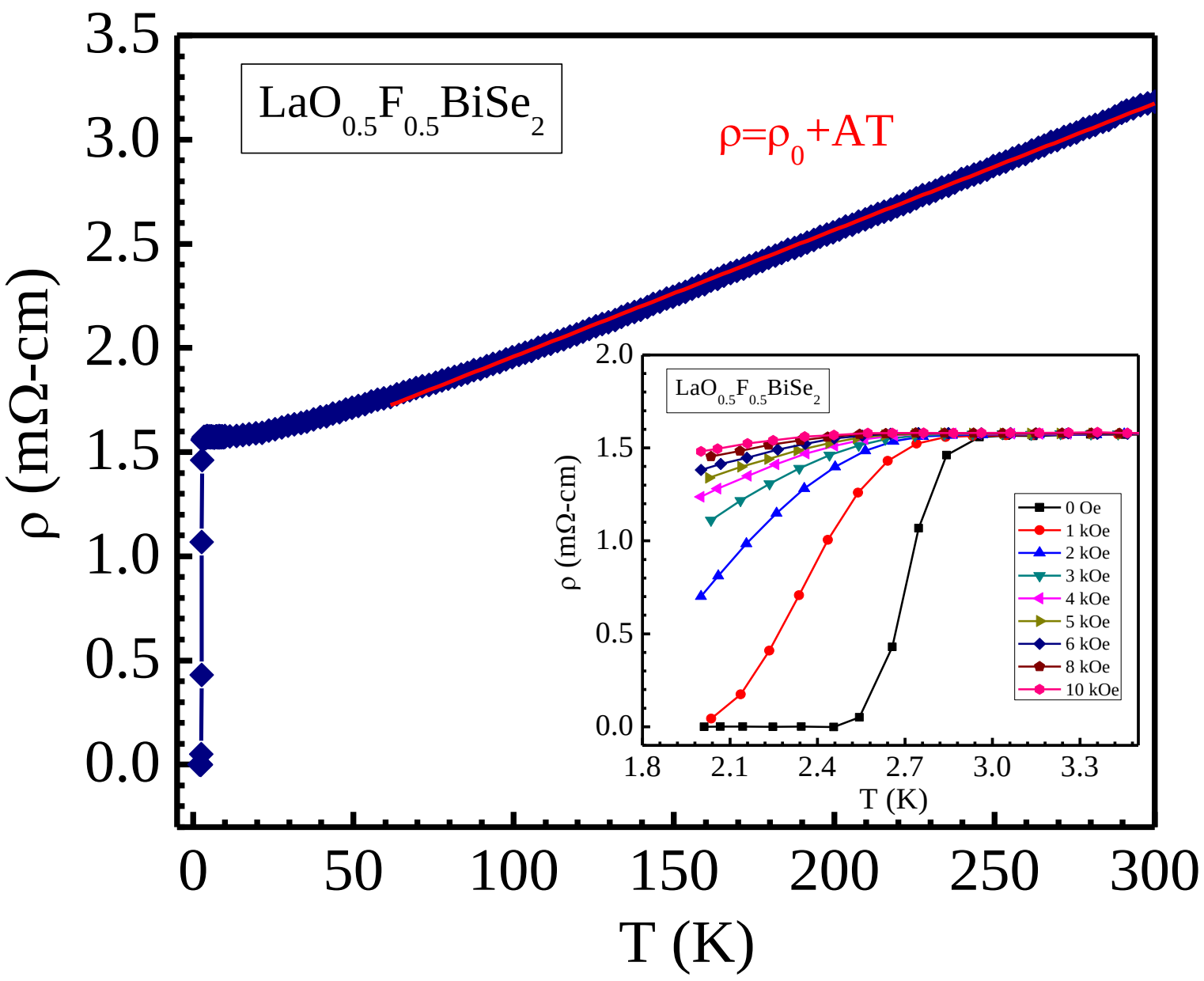

Fig. 4

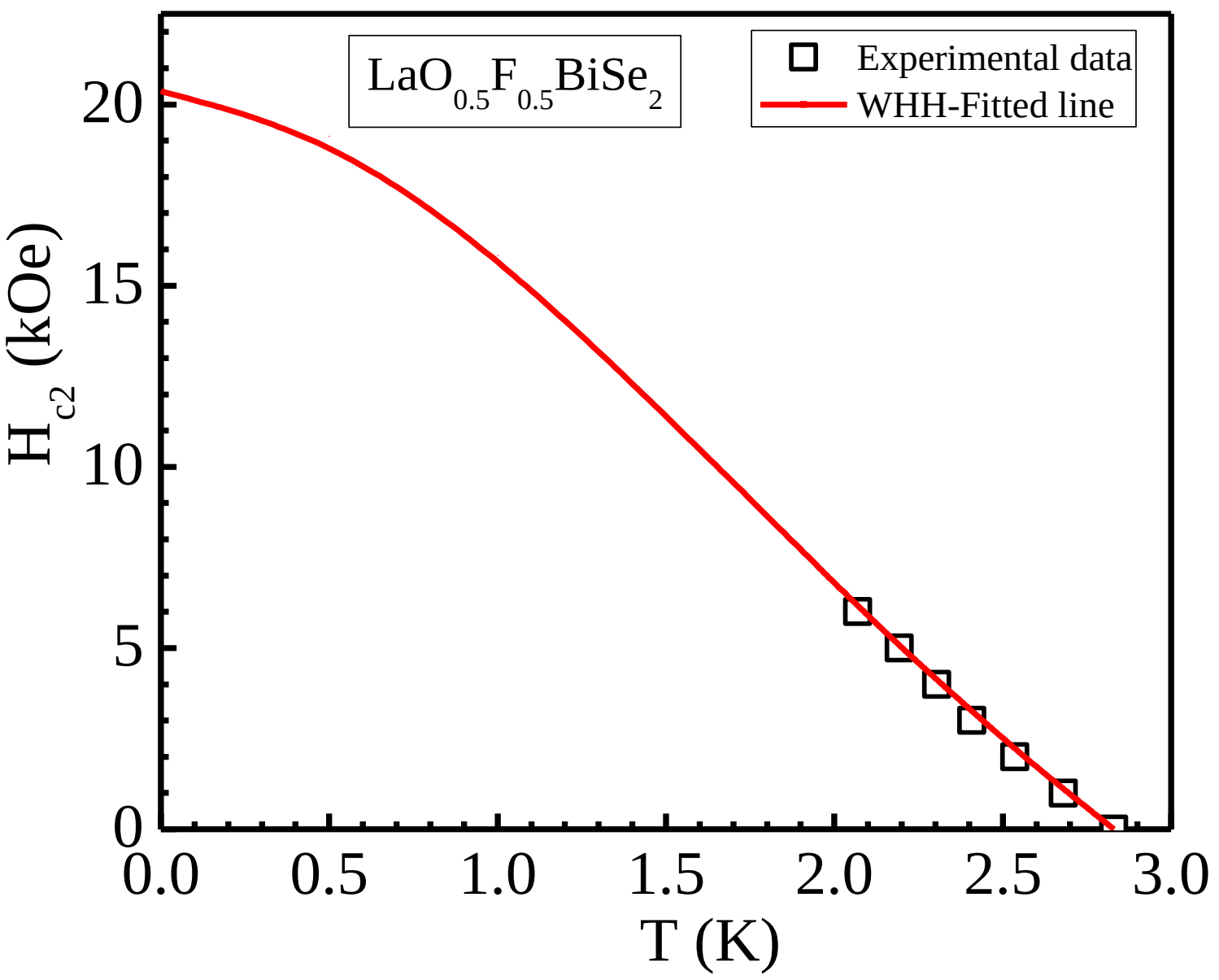

Fig.5

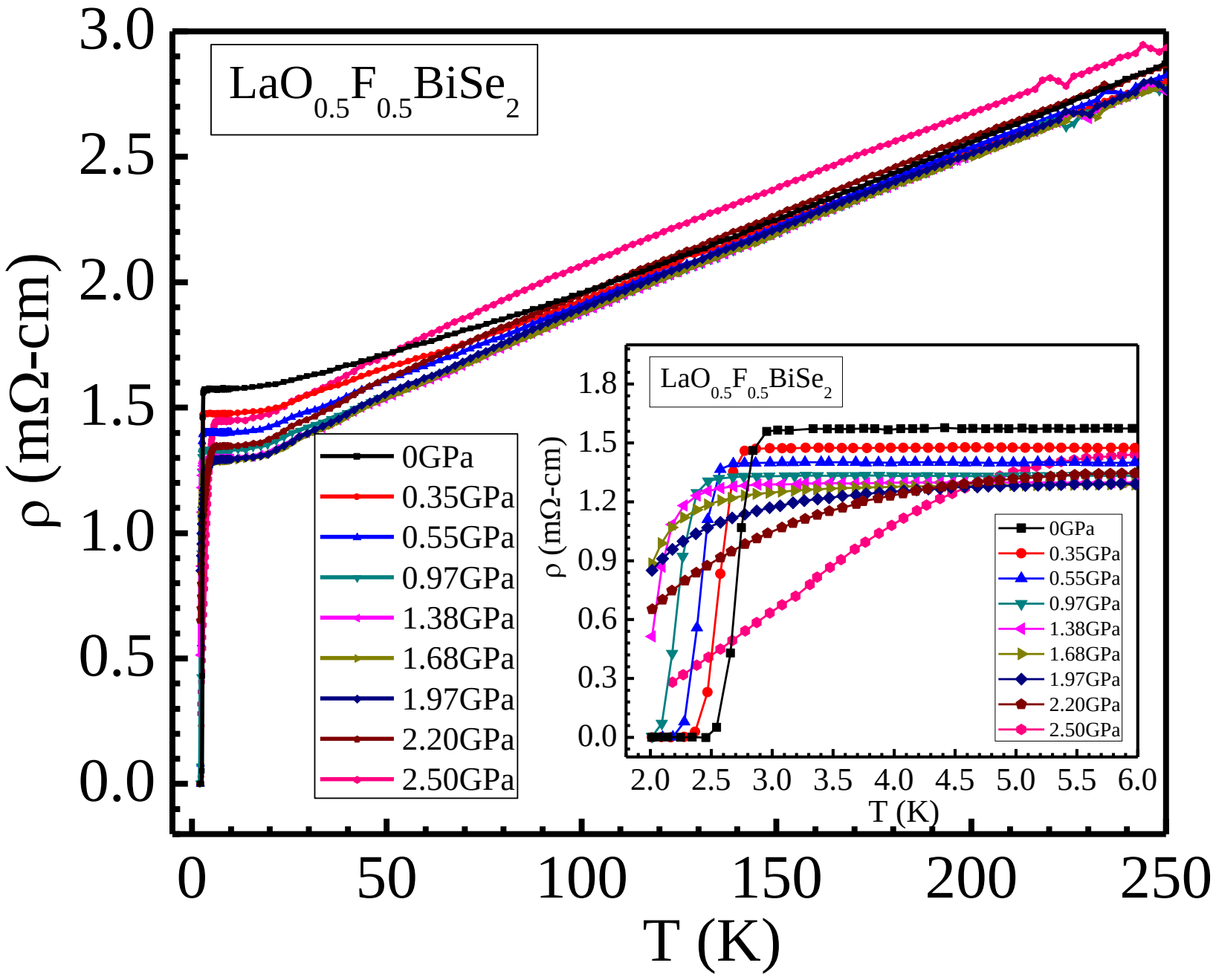